\documentclass[sigconf,screen=true]{acmart}

\usepackage[utf8]{inputenc}
\usepackage[T1]{fontenc}
\usepackage{microtype}
\usepackage[a-1b]{pdfx}   
\usepackage{graphicx}

\usepackage{amssymb, amsmath}
\usepackage{multirow}
\usepackage{framed}
\usepackage{listings}
\usepackage{booktabs} 
\graphicspath{ {./Images/} }
\newcommand{\greg}[1]{{\color{blue}\sc Greg: #1}}
\newcommand{\afonso}[1]{{\color{red}\sc Afonso: #1}}

\setcopyright{acmcopyright}
\acmPrice{15.00}
\acmDOI{10.1145/3472675.3473974}
\acmYear{2021}
\copyrightyear{2021}
\acmSubmissionID{fsews21toraclemain-id1-p}
\acmISBN{978-1-4503-8626-5/21/08}
\acmConference[TORACLE '21]{Proceedings of the 1st International Workshop on Test Oracles}{August 24, 2021}{Athens, Greece}
\acmBooktitle{Proceedings of the 1st International Workshop on Test Oracles (TORACLE '21), August 24, 2021, Athens, Greece}

\begin{document}

\title{Using Machine Learning to Generate Test Oracles: A Systematic Literature Review}

\author{Afonso Fontes}
\email{afonso.fontes@chalmers.se}
\affiliation{%
  \institution{Chalmers and the University of Gothenburg}
  \city{Gothenburg}
  \country{Sweden}
}

\author{Gregory Gay}
\orcid{0000-0001-6794-9585}
\email{greg@greggay.com}
\affiliation{%
  \institution{Chalmers and the University of Gothenburg}
  \city{Gothenburg}
  \country{Sweden}
}


\renewcommand{\shortauthors}{Afonso Fontes and Gregory Gay}

\begin{abstract}
Machine learning may enable the automated generation of test oracles. We have characterized emerging research in this area through a systematic literature review examining oracle types, researcher goals, the ML techniques applied, how the generation process was assessed, and the open research challenges in this emerging field. 

Based on 22 relevant studies, we observed that ML algorithms generated test verdict, metamorphic relation, and---most commonly---expected output oracles. Almost all studies employ a supervised or semi-supervised approach, trained on labeled system executions or code metadata---including neural networks, support vector machines, adaptive boosting, and decision trees. Oracles are evaluated using the mutation score, correct classifications, accuracy, and ROC. Work-to-date show great promise, but there are significant open challenges regarding the requirements imposed on training data, the complexity of modeled functions, the ML algorithms employed---and how they are applied---the benchmarks used by researchers, and replicability of the studies. We hope that our findings will serve as a roadmap and inspiration for researchers in this field.
\end{abstract}



\maketitle

\section{Introduction}\label{sec:intro}

\textit{Software testing} is invaluable in ensuring the reliability of the software that powers our society~\cite{STintro2011}. It is also notoriously difficult and expensive, with severe consequences for productivity, the environment, and human life if not conducted properly~\cite{Gay15:risks}. New tools and methodologies are needed to control that cost without reducing the quality of the testing process. Automation has a critical role to play in this effort by controlling testing costs and focusing developer attention on important tasks~\cite{Almasi17:IndustrialEval,Orso14:STR}. 

Consider test creation, an effort-intensive task that requires the selection of sequences of program input and \textit{oracles} that judge the correctness of the resulting execution~\cite{Harman13:oraclesurvey}. Automated test oracle creation is a topic of particular interest---and has earned the title ``the test oracle problem''~\cite{Harman13:oraclesurvey}. In current practice, oracles are often test-specific and require dedicated human effort to create. Advances have been made, but \textit{the test oracle problem remains unsolved}. If oracle creation could be even partially automated, developers' effort and cost savings could be immense. 

Advances in the field of machine learning (ML) have shown that algorithms can match or surpass human performance across many problem domains~\cite{MLintro1}. Machine learning has been used to advance the state-of-the-art in virtually every field. Automated test generation is no exception. We are interested in understanding and characterizing emerging research around the use of ML to generate or to support the creation of test oracles. 
Specifically, we are interested in understanding the types of oracles generated, the researchers' goals using ML, which specific ML techniques were applied, how such techniques were trained and validated, and how the success of the generation process was assessed. We also seek to identify limitations that must be overcome and open research challenges in this emerging field. 

To that end, we have performed a systematic literature review. Following a search of relevant databases and a rigorous filtering process, we have gathered a sample of 22 relevant studies. We have examined each study, gathering the data needed to answer our research questions. The findings of this study include: 
\begin{itemize}
\setlength{\itemsep}{1pt}
  \setlength{\parskip}{0pt}
  \setlength{\parsep}{0pt} 
    \item ML has been used to generate test verdict (18\%), metamorphic relation (27\%), and expected output (55\%) oracles.
    \item ML algorithms train predictive models that serve either as a stand-in for an existing test oracle---predicting a test verdict---or as a way to learn information about a function---either the expected output or metamorphic relations---that can be used as part of issuing a verdict.
    \item Almost all studies (96\%) employ supervised ML, trained on labeled system execution logs or source code metadata and validated based on the accuracy of the trained model.
    \item 59\% of the approaches employed  neural network (NN)---including Backpropagation NNs, Multilayer Perceptrons, RBF NNs, probabilistic NNs, and Deep NNs. 23\% of approaches adopted support vector machines. 5\% adopted decision trees, and another 5\% adopted adaptive boosting. The remaining 5\% did not specify a technique.
    \item Results were most often evaluated using the mutation score (55\%), followed by number of correct classifications (18\%), classification accuracy (18\%), and ROC (5\%). One study did not perform evaluation.
    \item The sampled studies show great promise, but there are still significant limitations and open challenges:
    \begin{itemize}
        \item Oracle generation is limited by the required quantity, quality, and content of training data. Assembling training data may require significant human effort. Models should be retrained over time.
        \item Applied techniques may be insufficient for modeling complex functions with many possible outputs. Varying degrees of output abstraction should be explored. Deep learning and ensemble techniques, as well as hyperparameter tuning, should be explored.
        \item Research is limited by overuse of toy examples, the lack of common benchmarks, and the inavailability of code and data. A benchmark should be created for evaluating oracle research, and researchers should be encouraged to provide replication packages and open code.
    \end{itemize}
\end{itemize}
\noindent Our study is the first to summarize this emerging research field. We hope that our findings will serve as a roadmap and inspiration for researchers interested in automated oracle generation.

\section{Background and Related Work}\label{sec:background}

\smallskip\noindent\textbf{Testing and Test Oracles:} Before complex software can be trusted, it is important to verify that the code is functioning as intended. Verification is often performed through the process of \textit{testing}---the application of \textit{input} to the system, and analysis of the resulting \textit{output}, to identify visible failures or other unexpected behaviors~\cite{STintro2011}.

During testing, a \textit{test suite} containing one or more \textit{test cases} is applied to the SUT. A test case consists of a \textit{test sequence (or procedure)}--a series of interactions with the SUT--with \textit{test input} applied to some component of the SUT. Input can range from a method call, to an API call, to an action taken within a graphical interface, depending on the granularity of the testing effort. Then, the test case will validate the output of the called components against a set of encoded expectations---the \textit{test oracle}---to determine whether the test passes or fails~\cite{STintro2011}. An oracle can be a predefined specification---encoded in a form usable by the test case---the output of another program, a past version of the SUT, or a model, or even manual inspection performed by humans. Most commonly, the oracle is formulated as a series of assertions on the values of output and stateful attributes~\cite{Harman13:oraclesurvey}. 

\begin{figure}[!t]
	\centering
	\begin{lstlisting}[language=Java,basicstyle={\scriptsize\ttfamily}]
	@Test
	public void testPrintMessage() {
	    String str = "Test Message";
	    TransformCase tCase = new TransformCase(str);
	    String upperCaseStr = str.toUpperCase();
	    assertEquals(upperCaseStr, tCase.getText());
	}
	\end{lstlisting} \vspace{-10pt}
	\caption{Example of a unit test. The \texttt{assertEquals} statement is an oracle, comparing the expected and actual output.}
	\label{fig:testcase} \vspace{-10pt}
\end{figure}

An example unit test is shown in Figure~\ref{fig:testcase}. The test passes a string to the constructor of the \texttt{TransformCase} class, then calls its \texttt{getText()} method to transform the string to upper-case. An assertion is used as an oracle to check whether the output is an upper-case version of the provided string. 

\smallskip\noindent\textbf{Machine Learning:} ML approaches construct models from observed data---and the structure of that data---to make decisions~\cite{MLbook2020}. Instead of being explicitly programmed with a set of instructions like in traditional software, ML algorithms ``learn'' from observations using statistical analyses, facilitating the automation of decision-making processes.  Learning begins with the search for patterns in a given dataset and, depending on the algorithm employed, may improve through new interactions over time. 

ML approaches largely fall into three categories: supervised, unsupervised, and reinforcement learning~\cite{MLbook2020}. In supervised learning, algorithms use previously labeled ``training'' data to infer a model that makes predictions about newly encountered data. In contrast to supervised methods, unsupervised algorithms do not make use of previously labeled data. Instead, approaches identify patterns based on the similarities and differences between data items. Rather than labeling items, unsupervised approaches are often used to cluster data and detect anomalies. Reinforcement learning algorithms select actions given their estimation of their ability to achieve some in-built goal, using feedback on the effect of the actions taken to improve their estimation of how to maximize achievement of this goal~\cite{Sutton2018}. Such algorithms are often the basis of automated processes, such as game bots or autonomous driving.

Recent ``deep learning'' (DL) approaches---often supervised--can make complex and highly accurate inferences from massive datasets that would be impossible in traditional ML approaches. This is because DL has an architecture inspired by organic neural networks that attempts to mimic how the human brain works~\cite{DLbook2016} using nonlinear processing layers where one layer's output serves as the successive layer's input. Deep learning requires a computationally intense training process and larger quantities of data than traditional supervised ML, but can learn highly accurate models, extract features and relationships from data automatically, and potentially apply models across applications.

\smallskip\noindent\textbf{Related Work:} To date, we are aware of no other systematic literature reviews dedicated to the use of ML to generate test oracles. However, there are secondary studies that cover overlapping topics. Most relevant is the survey on test oracles by Barr et al.~\cite{Harman13:oraclesurvey}. Their survey thoroughly summarizes research on test oracles up to 2014. They divide test oracles into four broad types, including those specified by human testers, those derived automatically from development artifacts, those that reflect implicit properties of all programs, and those that rely on a human-in-the-loop to judge test results. Approaches based on ML belong to the ``derived'' category, as they learn automatically from project artifacts to replace or augment human-written oracles. They discuss early approaches to using ML to derive oracles.

Durelli et al. performed a systematic mapping study on the application of ML to software testing~\cite{surveyMLinTesting2019}. Their scope is broader, but they do note that ML has been applied to support test oracle construction. They find that supervised learning is the most-used family of ML techniques overall software testing topics and that Artificial Neural Networks are the most used algorithm. 

Our study differs from the above through its focus specifically on the use of ML in oracle generation. This focus allows detailed analysis of this research area that is absent from broader surveys and mapping studies. Our study is also able to reflect more recent research than that covered in older studies.

\section{Methodology}\label{sec:methodology}

\begin{table*}[!t]
\centering
\scriptsize
\vspace{2mm}
\caption{List of research questions, along with motivation for answering the question.}
\label{tab:RQ} \vspace{-10pt}
\begin{tabular}{lll} 
\hline
\textbf{\textit{ID}} & \textbf{Research Question} & \textbf{Objective}   \\ 
\hline
\textit{\textbf{ RQ1} } & Which oracle types have been generated using ML? & \begin{tabular}[c]{@{}l@{}}Highlights test oracle types (e.g., information used to issue verdicts) targeted for ML-enhanced \\ oracle generation.\end{tabular} \\ \hline
\textit{\textbf{ RQ2} } & What is the goal of using machine learning as part of oracle generation?  & \begin{tabular}[c]{@{}l@{}}To understand the reasons for applying ML techniques to perform or enhance oracle generation \\ (e.g., potential benefits, expected result).\end{tabular}  \\ \hline
\textit{\textbf{ RQ3} } & How was machine learning integrated into the process of oracle generation? & \begin{tabular}[c]{@{}l@{}}Identifies how the ML technique was applied as part of the process of oracle generation, and \\ specify its training and validation steps. \end{tabular} \\ \hline
\textit{\textbf{ RQ4} } & Which ML techniques were used to perform or enhance oracle generation? & \begin{tabular}[c]{@{}l@{}}Identify specific ML techniques used in the process, including type, learning method, and  \\ selection mechanisms. \end{tabular} \\ \hline
\textit{\textbf{ RQ5} } & How is the oracle generation process evaluated?  & \begin{tabular}[c]{@{}l@{}}Describe the evaluation of the oracle generation process, highlighting any artifacts \\ (programs or datasets) they relied on. \end{tabular}   \\ \hline
\textit{\textbf{ RQ6} } & What are limitations and open challenges in ML-based oracle generation?  & \begin{tabular}[c]{@{}l@{}}Highlights the limitations of oracle generation, such as data dependency, accuracy, or \\ training time, and challenges that must be overcome to apply oracle generation in the field.\end{tabular}  \\ \hline
\end{tabular} \vspace{-5pt}
\end{table*}

Our concern in this work is to understand how researchers have used machine learning (ML) to perform, or otherwise enhance, automated test oracle generation. We have investigated contributions to the literature related to this topic and seek to understand their methodology, results, and insights. To achieve this task, it is necessary to carry out a secondary study---specifically a Systematic Literature Review (SLR)~\cite{Kitchenham07:Guidelines}. This section describes how we conducted our SLR. 

We are interested in assessing the \textit{effect} of integrating ML into the oracle generation process, understanding the \textit{adoption} of these techniques---how and why they are being integrated, and which specific techniques are being applied, and identifying the potential \textit{impact} and \textit{risks} of this integration. Table~\ref{tab:RQ} lists the research questions we are interested in answering, briefly defines why those questions are important, and lists the properties extracted from primary studies to answer them (defined in Section \ref{sec:collection}).

Questions 1-3 allow us to understand how ML techniques have enhanced oracle generation, why they were applied, and which specific oracle types were targeted. \textbf{RQ2} is motivational, covering the authors' primary objectives. In contrast, \textbf{RQ3} expressly is a technical question, examining the specific roles of the included ML techniques, as well as its training and validation processes.

\textbf{RQ4} examines which ML techniques were used to perform the generation task, as well as \textit{why} that specific method was adopted, if the authors provide such information. \textbf{RQ5} focuses on how the oracle generation approach is evaluated. Finally, \textbf{RQ6} aims to cover the limitations of the proposed approaches, open issues, and insights that we have uncovered in this area.
To answer these questions, we have done the following:
\begin{enumerate}
\setlength{\itemsep}{1pt}
  \setlength{\parskip}{0pt}
  \setlength{\parsep}{0pt}  
    \item Formed a list of studies (Section~\ref{sec:selection}).
    \item Filtered this list for relevance (Section~\ref{sec:filtering}).
    \item Extracted data from each study, guided by a set of properties of interest (Section~\ref{sec:collection}).
    \item Identified trends in the extracted data (Section~\ref{sec:results}).
\end{enumerate}

\subsection{Initial Study Selection}\label{sec:selection}

To locate studies for consideration, a search was conducted using four databases: IEEE Xplore, ACM Digital Library, Science Direct, and Scopus. We created a search string to narrow the results by combining terms of interest regarding automated test generation and machine learning. Note that our search was purposefully broad, intended to capture studies using ML to enhance both input and oracle generation.  This approach allowed us to capture a wide range of studies, including those that a narrow search would miss. We then filtered the pool for relevancy. Each database uses a different search engine, and the search options and search formulation slightly vary between them. In general, the search string used was:
\begin{center}
\textit{(``test case generation'' OR ``test generation'' OR ``test oracle'' OR ``test input'') AND (``machine learning'' OR ``reinforcement learning'' OR ``deep learning'' OR ``neural network'')}
\end{center}
\noindent These keywords are not guaranteed to capture all existing relevant articles. However, they are designed to capture a sufficiently wide sample to answer our research questions. Specifically, we combine terms related to test case generation---including specific test components---and terms related to machine learning---including common technologies.

Our focus is specifically on the use of ML in oracle generation, not on any form of automated oracle generation. To obtain a representative sample, we have selected ML-related terms that we expect will capture a wide range of studies. These terms may omit some oracle generation techniques that could be in-scope, but allow us to obtain a representative sample while controlling the number of studies that require manual inspection.

\begin{figure}[!t]
\centering
\includegraphics[width=0.8\columnwidth]{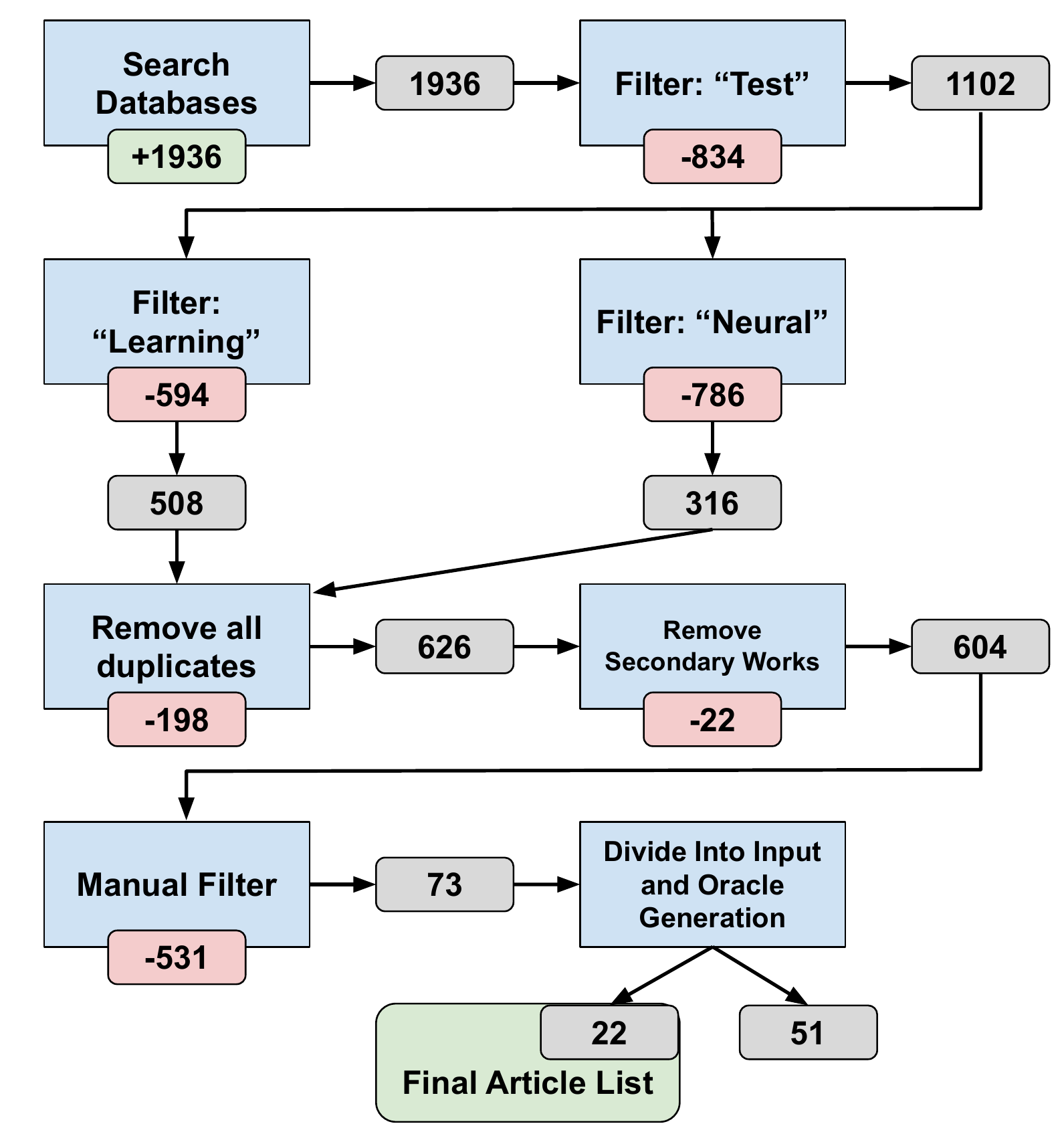} \vspace{-5pt}
\caption{Steps taken to determine the final list of studies.}
\label{fig:filtersEntries} \vspace{-15pt}
\end{figure}

Before exporting the results, we applied an initial filter to the results using the advanced search option in each database, which consists of the following selection criteria: (a) published studies in conferences and journals (excluding grey literature such as pre-prints, technical reports, abstracts, editorials, and book chapters); (b) studies published before November 2020 (when we conducted the search); (c) studies written in the English language. After exporting all results, a total of 1936 studies were identified. This is shown as the first step in Figure~\ref{fig:filtersEntries}. 

To evaluate the search string's effectiveness, we conducted a three-step verification process. First, we randomly sampled ten entries from the 73 studies that remained following the manual filtering. Then we looked in each article for ten citations that might also be in scope, resulting in a list of 100 citations. We checked whether the search string also retrieved the citations in the list, and all 100 were retrieved by the string (pre-filtering). Although this is a small sample, it indicates the robustness of the search string. 

After the search, the next step was to identify whether secondary studies already existed on this topic. If so, the need for this SLR would be reduced. We found no previous secondary studies focusing specifically on ML-based oracle generation. However, we identified a small number of related studies. These are discussed in Section~\ref{sec:background}.

\begin{table*}[!t]
\centering
\scriptsize
\caption{List of properties used to answer the research questions. For each property, we include a name, the research questions the property is associated with, and a short description.}
\label{tab:props}
\vspace{-10pt}
\begin{tabular}{llll} 
\hline
\textit{\textbf{ID}}   & \textbf{Property Name} & \textbf{RQ} & \textbf{Description}  \\  \hline
\textit{\textbf{P1}}   & Test Oracle Type  & RQ1, RQ2  & The specific type of oracle focused on by the approach. It helps to categorize the studies, enabling comparison between contributions. \\ \hline
\textit{\textbf{P2}}   & Proposed Research  & RQ2   & A short description of the approach proposed or research performed.  \\ \hline
\textit{\textbf{P3}}   & Hypotheses and Results  & RQ1, RQ3    & Highlights the differences between expectations and conclusions of the proposed approach. \\ \hline
\textit{\textbf{P4}}   & ML Integration & RQ3   & \begin{tabular}[c]{@{}l@{}}Covers how ML techniques have been integrated into the oracle generation process. It is essential to understand what aspects of \\ generation are handled or supported by ML. \end{tabular}   \\ \hline
\textit{\textbf{P5}}   & ML Technique Applied  & RQ4  & Name, type, and description of the ML technique used in the study.  \\ \hline
\textit{\textbf{P6}}   & Reasons for Using the Technique & RQ4   & The reasons stated by the authors for choosing this particular ML technique. \\ \hline
\textit{\textbf{P7}}   & ML Training Process  & RQ4  & \begin{tabular}[c]{@{}l@{}}How the approach was trained, including the specific data sets or artifacts used to perform this training. Helps us understand how \\ contributions  could be replicated or extended. \end{tabular}  \\ \hline
\textit{\textbf{P8}}   & External Tools or Libraries Used & RQ4 & External tools or libraries used to implement the ML technique. \\ \hline
\textit{\textbf{P9}}   & ML Objective and Validation Process & RQ4, RQ5    & \begin{tabular}[c]{@{}l@{}}The objective of the ML technique (i.e., validation metric), and how it is validated, including data, artifacts, and \\ metrics used (if any). \end{tabular}   \\ \hline
\textit{\textbf{P10}}  & Oracle Creation Evaluation Process & RQ5  & \begin{tabular}[c]{@{}l@{}}Covers how the ML-enhanced oracle generation process, as a whole, is evaluated (i.e., how successful are the generated oracles \\ at detecting faults or meeting some other testing goal?). Allows understanding of the effects of ML on improving the testing process.
\end{tabular}  \\ \hline
\textit{\textbf{P11}}  & Potential Research Threats  & RQ6 & Notes on the threats to validity that could impact each study.  \\ \hline
\textit{\textbf{P12}}  & Strengths and Limitations  & RQ6 & Used to understand the general strengths and limitations of enhancing oracle creation with ML. \\ \hline
\textit{\textbf{P13}}  & Future Work & RQ6 & Any future extensions proposed by the authors, with a particular focus on those that could overcome identified limitations. \\ \hline
\end{tabular} \vspace{-5pt}
\end{table*}

\begin{figure}[!t]
\centering
\includegraphics[width=0.95\columnwidth]{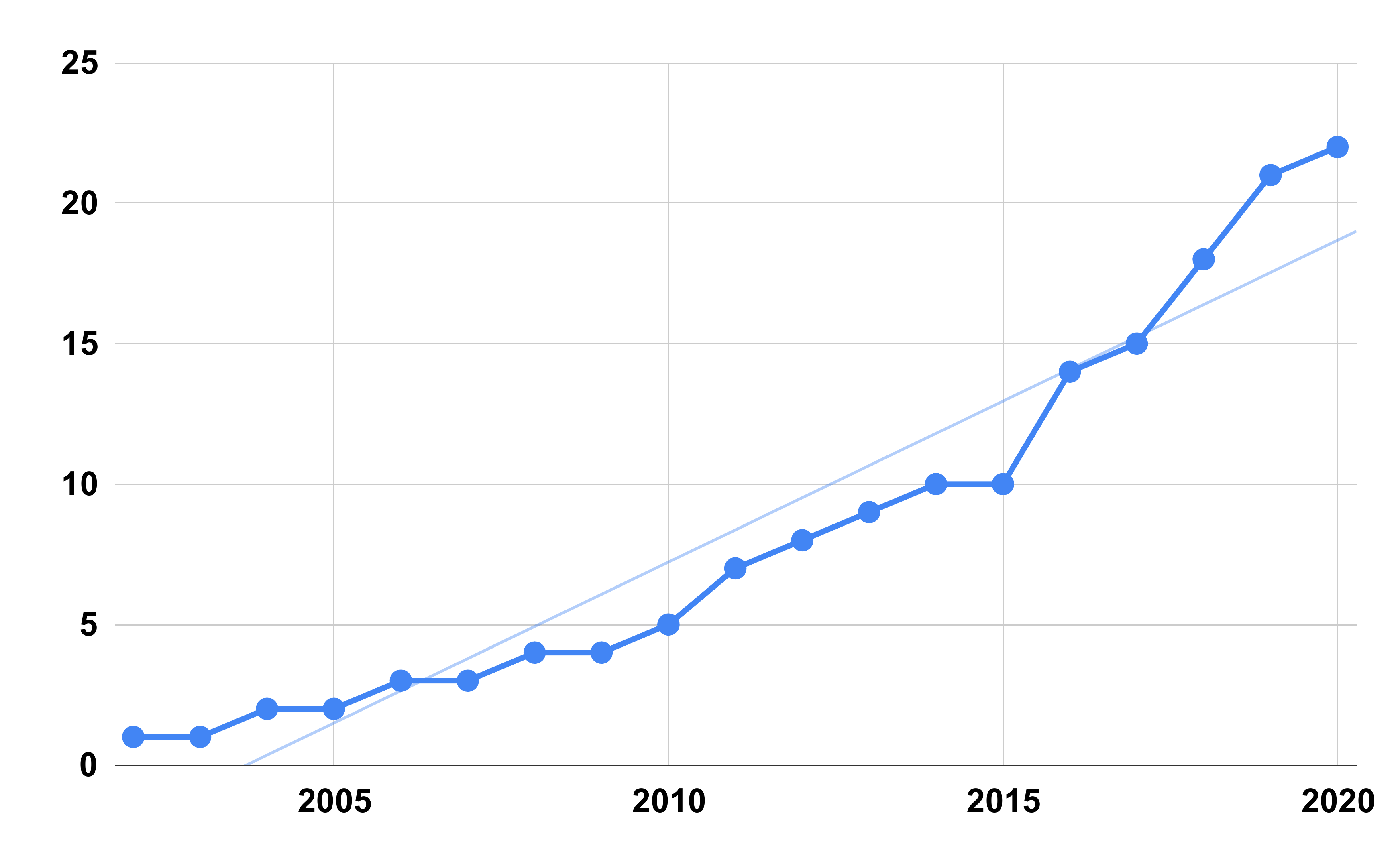} \vspace{-15pt}
\caption{Growth in the number of publications in ML-based oracle generation from 2002-2020.}
\label{fig:pubyear} \vspace{-15pt}
\end{figure}

\subsection{Selection Filtering}\label{sec:filtering}

The initial search resulted in 1,936 studies. It is unlikely that all would be relevant. Therefore, we applied a series of filtering steps to obtain a focused list. Figure~\ref{fig:filtersEntries} presents the filtering process and the number of entries after applying each filter. The tag in the center of box 1 represents the 1,936 studies exported from the search and added to the list. The tags in the other boxes represent the number of entries removed in that particular step. The numbers between boxes show the total number of articles that resulted after applying the previous step. Finally, the highlighted box at the end shows the final number of studies used to answer our research questions.

To ensure relevancy, we used a set of keywords to filter the list. We first searched the title and abstract of each study for the keyword ``test''. This step removed 834 articles. We then searched the resulting list for either ``learning'' or ``neural''---representing the application of machine learning. Every article from IEEE Xplore and Scopus passed these filters. However, the number of articles from the ACM Digital Library and Science Direct was significantly reduced. We merged the filtered lists for both keywords. Some studies contained both keywords in the title or abstract. To remove these, as well as any studies that were returned by multiple databases, we removed all duplicate entries, which resulted in 626 remaining studies. We then removed 22 secondary studies, leaving 604 studies.

We examined the remaining studies manually, removing all not in scope following an inspection of the title and abstract. We removed any studies not related to software test generation or that do not apply ML during the test generation process (i.e., the ML element is related to a particular activity such as test suite reduction). This determination was made by first reading the abstract of the paper, then the introduction, then the full paper, until a clear determination could be made of the relevancy of the study. Both authors independently inspected studies during this step to prevent the accidental removal of relevant studies. In cases of disagreement, the authors discussed and came to a conclusion. This left 73 studies. Finally, we divided these studies into those related to input or oracle generation. This step resulted in a final total of 22 studies related to oracle generation for consideration. 

Figure~\ref{fig:pubyear} shows the rate of growth in this emerging research area. The first study in our sample was published in 2002 and the most recent in 2020. Interest in this topic is growing with the emergence of new and more powerful ML approaches, with over half of the studies having been published since 2016. 

\subsection{Data Extraction}\label{sec:collection}

To answer the questions listed in Table~\ref{tab:RQ}, we have examined each study. We have focused on a set of key properties, identified in Table~\ref{tab:props}. Each property listed in the table is briefly defined and is associated with the research questions that it will help answer. In many cases, several properties are collectively used to answer a RQ. For example, the answer to RQ2, which aims to cover the goals of using ML as part of the automated test generation process, can be extracted from property P2 in many cases. However, P1 is related because it provides context to the research and the particular type of test oracle may dictate how ML is applied. Each property is important in capturing the essential details of the study and how it contributes to answering our RQs. 

In reported experiments, the proposed approach either exceeded or failed to meet the initial hypotheses. This is covered by the third property, P3, which could lead to or be part of the answer for RQ1 and RQ3. The fourth property targets RQ3 and notes how the adopted ML technique is integrated into the testing process. To understand how ML techniques can enhance automated test generation, it is important to understand which techniques are applied as well as the motivation behind adopting a specific technique. These aspects are covered by P5 and P6, which are used to answer RQ4. We also note whether the project analyzed is new or the continuation of prior research as part of collecting data for these properties.

The following three properties focus on understanding the application of ML in the study, including a partial assessment of the potential to replicate the research, by covering core characteristics of the ML technique---the training process (P7), external tools used to implement the technique (P8), and the validation process (P9). P7 focuses on the datasets or other information sources used to train the learning technique. Our primary focus with P8 is to cover how external tools, environments, or ML libraries---such as TensorFlow or Keras---are used to train, build, or execute the ML technique. The combination of properties P7, P8, and P9 will answer RQ4, which examines how the ML technique is trained, validated, and assessed as part of its integration. RQ5 examines how the entire oracle generation process is evaluated. P10 is primarily used to answer this research question. However, P9 may also help answer this question.

Research question RQ6 covers open challenges. Properties P11-P13 contribute to answering this question, including limitations and threats to validity---either disclosed by the authors or inferred from our analysis---and future work. 

Data extraction was performed primarily by the first author of this study. However, to ensure the accuracy of the extraction process, the second author performed an independent extraction for a randomly-chosen sample of the studies. We compared our findings, and found that we had near-total agreement.  

\section{Results and Discussion}\label{sec:results}

We divide the examination of the results as follows: the types of oracles generated using ML and why ML was applied (RQ1-2, Section~\ref{sec:problems}), how ML was applied in the examined studies (RQ3-5, Section~\ref{sec:use}), and the limitations and open challenges in this emerging research field (RQ6, Section~\ref{sec:challenges}). 

\subsection{Test Oracle Types and Motivation}\label{sec:problems}

\begin{figure}[!t]
\centering
\vspace{2mm}
\includegraphics[width=0.95\columnwidth]{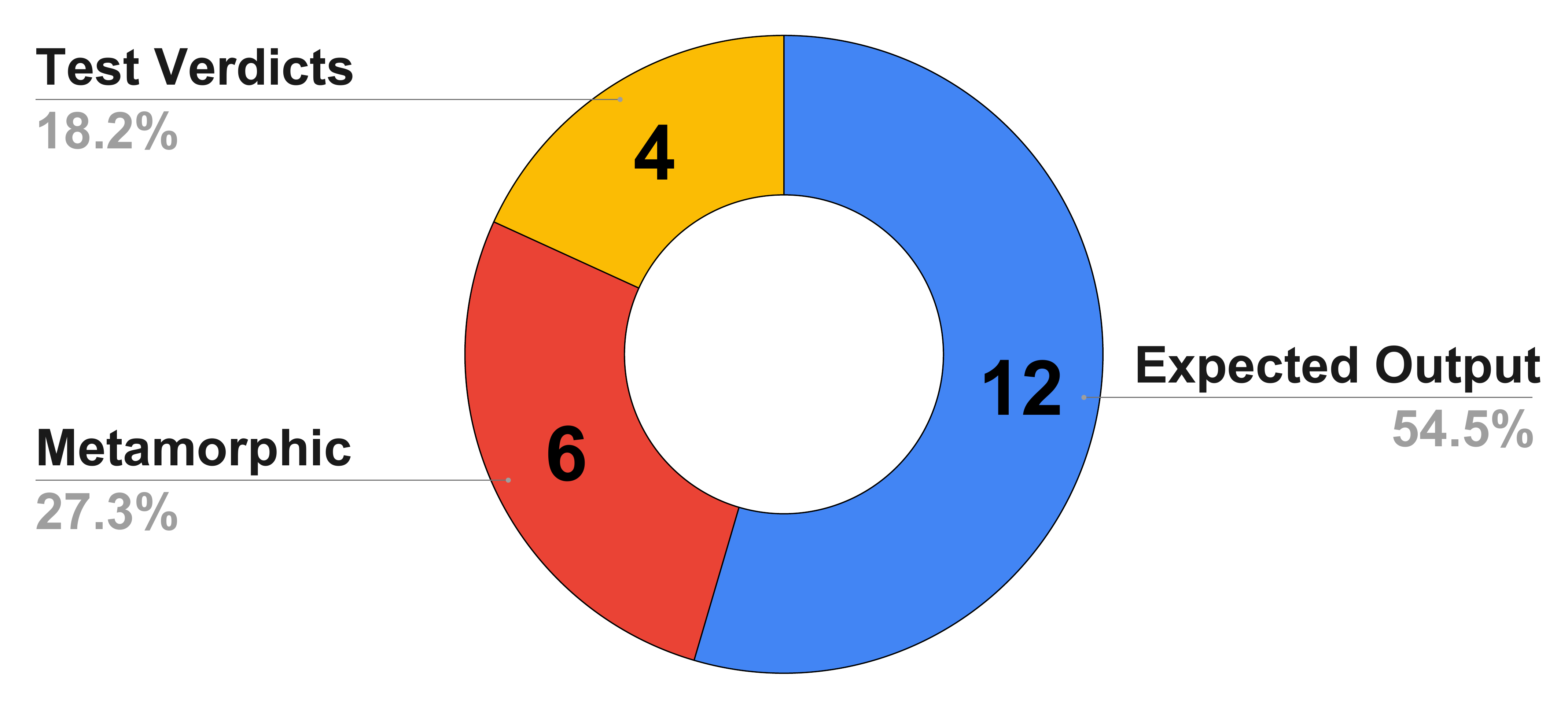} \vspace{-15pt}
\caption{The types of oracles generated, and the number of studies where this type of oracle is generated.}
\label{fig:articleDivision} \vspace{-10pt}
\end{figure}

\begin{table*}[!t]
\scriptsize
\centering
\vspace{2mm}
\caption{Data on the sampled studies, including the type of ML approach, specific ML technique, training data used, the targeted goal of the ML approach, how the approach was evaluated, and the type of application used in the evaluation.}
\label{tab:oracle} \vspace{-10pt}
\begin{tabular}{lllllllll}
\hline
\textbf{Ref}  & \textbf{Year} & \textbf{Oracle Type}  & \textbf{ML Approach} & \textbf{Technique} & \textbf{Training Data} & \textbf{ML Objective}  & \textbf{Evaluation Metric} & \textbf{Evaluated On} \\ \hline
\cite{Braga2018}   & 2018  & Test Verdicts   & Supervised & Adaptive Boosting & System Executions & Regression  & Mutation Score  & Shopping Cart \\ \hline
\cite{Gholami2018}  & 2018  & Test Verdicts  & Supervised & Backpropagation NN & System Executions & Regression & Mutation Score & Embedded Software  \\ \hline
\cite{Makondo2016}  & 2016 & Test Verdicts   & Supervised  & Multilayer Perceptron & System Executions  & Regression  & Accuracy & User Creation\\ \hline
\cite{Shahamiri2010a} & 2010 & Test Verdicts   & Supervised   & Backpropagation NN   & System Executions      & Regression   & Mutation Score    & Student Registration        \\ \hline

\cite{Aggarwal2004}   & 2004 & Expected Output     & Supervised & Backpropagation NN    & System Executions    & Regression  & Correct Classifications & Triangle Classification \\ \hline
\cite{Ding2016} & 2016 & Expected Output    & Supervised & SVM  & System Executions      & Label Propagation  & Mutation Score  & Image Processing   \\ \hline
\cite{Jin2008} & 2008  & Expected Output    & Supervised & Backpropagation NN    & System Executions & Regression   & Correct Classifications  & Triangle Classification \\ \hline
\cite{Monsefi2019} & 2019   & Expected Output   & Supervised & Deep NN   & System Executions & Regression & Mutation Score   & Mathematical Functions  \\ \hline
\cite{Sangwan2011} & 2011   & Expected Output  & Supervised  & RBF NN & System Executions  & Regression  & Correct Classifications & Triangle Classification \\ \hline
\cite{Shahamiri2011}  & 2011 & Expected Output   & Supervised   & Multilayer Perceptron & System Executions  & Regression   & Mutation Score    & Insurance Application   \\ \hline
\cite{Shahamiri2012}  & 2012 & Expected Output   & Supervised   & Multilayer Perceptron & System Executions     & Regression & Mutation Score       & Insurance Application   \\ \hline
\cite{Singhal2016}    & 2016 & Expected Output    & Supervised   & Backpropagation NN + Cascade & System Executions      & Regression & Accuracy & Credit Analysis     \\ \hline
\cite{Vanmali2002}   & 2002 & Expected Output    & Supervised   & Not Specified  & System Executions      & Regression   & Mutation Score     & Credit Analysis   \\ \hline
\cite{Vineeta2014a}  & 2014 & Expected Output    & Supervised  & Backpropagation NN    & System Executions      & Regression     & Mutation Score         & Triangle Classification    \\ \hline
\cite{Ye2006a}   & 2006     & Expected Output    & Supervised   & Multilayer Perceptron & System Executions      & Regression   & Mutation Score &  Mathematical Functions  \\ \hline
\cite{Zhang2019d}  & 2019   & Expected Output   & Supervised   & Probabilistic NN & System Executions      & Regression                       & Correct Classifications & Prime, Triangle Class \\ \hline

\cite{Hardin2018}  & 2018   & Metamorphic  & Supervised & SVM  & Code Features & Label Propagation & Accuracy &   Various Functions \\ \hline
\cite{Hiremath2020}  & 2020 & Metamorphic    & Reinforcement & Not Specified  & System Executions      & Discovered Relations & Not Evaluated & Ocean Modeling    \\ \hline
\cite{Kanewala2013b} & 2013 & Metamorphic  & Supervised   & SVM, Decision Trees   & Code Features & Regression & Mutation Score & Various Functions \\ \hline
\cite{Kanewala2016}  & 2016 & Metamorphic     & Supervised    & SVM   & Code Features  & Regression & Mutation Score  & Various Functions\\ \hline
\cite{Nair2019}  & 2019 & Metamorphic         & Supervised   & SVM      & Code Features      & Label Propagation & ROC & Matrix Calculation   \\ \hline
\cite{Zhang2017}   & 2017   & Metamorphic           & Supervised  & RBF NN & Code Features     & Multi-label Regression & Accuracy & Various Functions\\ \hline

\end{tabular} \vspace{-10pt}
\end{table*}

Before examining which ML techniques have been integrated into oracle generation, or how they have been integrated, it is first crucial to understand \textit{why} they have been integrated. A test oracle is a broad, high-level concept---simply \textit{some} means to judge the correctness of the system given test input~\cite{Harman13:oraclesurvey}. Therefore, our first two research questions are intended to give an overview of the specific types of oracle that have been the focus of the collected studies (RQ1) and to provide motivation for why ML was applied as part of creating these oracles. Figure~\ref{fig:articleDivision} shows our results. Broadly, three types of oracles have been generated in the examined studies: 
\begin{itemize}
\setlength{\itemsep}{1pt}
  \setlength{\parskip}{0pt}
  \setlength{\parsep}{0pt}  
    \item \textbf{Test Verdicts:} The approach directly predicts the final test verdict, given provided input. For example, this type of oracle might directly issue a ``pass'' or ``fail'' verdict for the test case.
    \item \textbf{Expected Output:} The approach predicts specific system behavior that should result from applying the provided input~\cite{Gay15:oracleselection}. The predicted behavior can vary in its level of abstraction, from a concrete output to a broad \textit{class} of output---generally leaning more towards the abstract, given the challenges of making specific predictions for complex systems.
    \item \textbf{Metamorphic Relations:} A metamorphic relation is a necessary property of a function, relating input to the output produced~\cite{Hardin2018}. For example, a metamorphic relation for a $sine$ function is $sin(x) = sin(\pi - x)$. Such relations allow us to infer expected results for different input values to a function, and violations of such properties identify potential faults. Approaches in this category attempt to learn metamorphic relations for new systems from provided data. 
\end{itemize}
\noindent Of the 22 collected studies, a majority---12 approaches---produce expected output oracles. Six produce metamorphic relations, and four produce direct test verdicts. 

The goal of ML is to automate or support a decision process. Given an observation, a ML technique can make a prediction. That prediction can either be the final decision to be made, or it can relate to a piece of information needed to make that decision. Test oracles follow a similar conceptual model. Test oracles consist of two core components---the oracle \textit{information}, or a set of facts used to issue the verdict on the test case, and the oracle \textit{procedure}, the actions taken to issue a verdict based on the embedded information and observations of system behavior~\cite{Richardson92:TestOracles}. Motivationally, we can see that ML offers a natural means to replace either the oracle information---which typically requires human effort to specify---or the oracle as a whole. Test verdict oracles perform the entire decision process, directly issuing a verdict. The other two oracle types, expected outputs and metamorphic relations, replace human specification of oracle information with a model that predicts that information instead. The procedure can then act on that prediction rather than relying on human-specified facts.

\begin{center}
\begin{framed}
\vspace{-5pt}
  \textbf{RQ1 (Oracle Types):} Machine Learning algorithms have been used to generate test verdict (18\%), metamorphic relation (27\%), and expected output (55\%) oracles.
  \vspace{-5pt}
\end{framed}
\end{center}

\begin{center}
\begin{framed}
\vspace{-5pt}
  \textbf{RQ2 (Goal of ML):} ML algorithms train models that serve either as a stand-in for a test oracle or to learn information about a function (e.g., expected output or metamorphic relations) that can be used as part of issuing a verdict.
  \vspace{-5pt}
\end{framed}
\end{center}

\subsection{Application of Machine Learning}\label{sec:use}

Table~\ref{tab:oracle} summarizes relevant data gathered from the 22 studies where ML was used to generate test oracles. Immediately, we can see that almost all approaches adopted a supervised approach, where a model is trained and used to make predictions about new input. Unsupervised and reinforcement learning (RL) have been used as part of input generation. These approaches may also be applicable as part of oracle generation---e.g., an oracle modeled as a RL agent could make predictions and get feedback on their accuracy, or an unsupervised clustering approach could be used as part of an oracle that detects anomalies. One study did propose the use of RL-like techniques as part of metamorphic relation generation. However, the focus has been firmly on supervised learning. 

The sampled studies train oracles using a set of previously-captured and labeled system executions or metadata about source code features. The model is then used to predict the correctness of new behaviors or to predict the type of behavior that will result from applying the input. We will discuss each oracle type in turn.

\smallskip\noindent\textbf{Test Verdicts:} All studies within this category applied a ML technique to associate patterns in the training data with the resulting test verdict (i.e., they trained a model for the purpose of regression). This approach enables the oracles generated to assert whether a test passes or fails without running the SUT. 

Makondo et al.~\cite{Makondo2016} utilize a Multilayer Perception (MLP) Neural Network (NN)---a basic NN, often constructed with a single hidden layer. Shahamiri et al.~\cite{Shahamiri2010a} and Gholami et al.~\cite{Gholami2018} utilized Feed-forward Backpropagation (BP) NNs to create their test oracles. A BP NN ``learns'' by reducing error rates by tuning the weights in each neuron after computing the error, making the model more stable. Braga et al.~\cite{Braga2018} use a classifier based on adaptive boosting.

Braga et al.~\cite{Braga2018} gather usage data from a shopping website by inserting several specific capture components into the site. The data then goes through a preprocessing step and then is finally used for training the ML. Shahamiri et al.~\cite{Shahamiri2010a} focus on a student registration-verifier application that checks whether a students' records satisfy the minimum requirements for enrollment. Gholami et al.~\cite{Gholami2018} focus on embedded systems in their evaluation. Makondo et al.~\cite{Makondo2016} examined a user creation function. Braga et al.~\cite{Braga2018}, Gholami et al.~\cite{Gholami2018} and Shahamiri et al.~\cite{Shahamiri2010a}) evaluate their approaches using the mutation score. They insert synthetic faults, and measure how many of these faults that the generated oracle can detect. Makondo et al.~\cite{Makondo2016} evaluate using the accuracy of the classification model.

\smallskip\noindent\textbf{Expected Output:} More than half of the studies generate expected output oracles. The approaches train on system executions, and then predict the output given a new input. Often, the level of detail of the output generated is constrained or abstracted to a small set of representative values, rather than attempting to predict highly specific output. For example, rather than yielding a specific integer for integer output, the approach might constrain the output to a limited set of representative values (classifications) and predict one of those values. Otherwise, evaluation is limited to code that issues output from an enumerated set of values. A common application is the ``triangle classification problem,'' also known as TRITYP~\cite{Aggarwal2004,Sangwan2011,Vineeta2014a, Zhang2019d,Sangwan2011}. The program receives three numbers representing the lengths of a triangle's sides and outputs a classification of the type of triangle as scalene, isosceles, equilateral, or not a triangle. This is a problem that can prove challenging given its branching behavior. However, it still has a limited set of output possibilities. This makes it a reasonable starting point for oracle generation. 

Zhang et al.~\cite{Zhang2019d} also model a function that judges whether an integer is prime or not. This is an even more straightforward application---a two-class classification problem. Shahamiri et al.~\cite{Shahamiri2011,Shahamiri2012} generate oracles for a car insurance application, while Singhal et al.~\cite{Singhal2016} and Vanmali et al.~\cite{Vanmali2002} generate oracles for a credit analysis at a bank. Ding et al.~\cite{Ding2016} generate oracles for an image processing function that classifies a type of cell from image sections. All of these applications produce output from an enumerated set of values, easing the difficulty of generating an oracle. 

Ye et al.~\cite{Ye2006a} and Monsefi et al.~\cite{Monsefi2019} generate oracles for functions with integer output. Some of the cases they examine have a limited range of produced outputs (e.g., a function that predicts the length of a route). Still, the remaining functions offer some indication that deep learning can model more complex functions or predict more detailed expected output.

Ding et al.~\cite{Ding2016} used a support vector machine (SVM) to perform label propagation. Label propagation is a semi-supervised learning technique, where a mixture of labeled and unlabeled training data is used to train the model, and the algorithm attempts to propagate labels from the labelled data to similar, unlabeled data. This can reduce the quantity of training data needed. 

The other approaches follow a more traditional supervised, regression-based learning process, and generally make use of different NNs. Four of the examined studies adopt a Backpropagation NN~\cite{Aggarwal2004, Jin2008, Singhal2016, Vineeta2014a}. Three other studies employ the Multilayer Perceptron technique~\cite{Shahamiri2011, Shahamiri2012, Ye2006a}. Sangwan et al. uses a Radial Basis Function (RBF) NN~\cite{Sangwan2011}. RBF is a specific activation function applied to the inputs of the network. Monsefi et. al~\cite{Monsefi2019} adopt a Deep NN, which has more input and output layers than a regular NN, with a fuzzy encoder + decoder. Finally, Zhang et. al adopt a probabilistic NN~\cite{Zhang2019d}.

In terms of evaluation, five of the studies are focused on the accuracy of the oracle in a set of cases where the ground truth is known---measuring the percentage of correct classifications~\cite{Zhang2019d,Aggarwal2004,Jin2008,Sangwan2011} or the accuracy of the model~\cite{Singhal2016}. The remaining seven used the mutation score as the evaluation metric~\cite{Ding2016,Monsefi2019,Shahamiri2011,Shahamiri2012,Vanmali2002,Vineeta2014a,Ye2006a}. 

\smallskip\noindent\textbf{Metamorphic Relations:} Six approaches generate metamorphic relations---properties of a function that explain how particular input links to  corresponding output~\cite{Hardin2018}. Such relations allow us to infer expected results for different input values to a function, and violations of such properties identify potential faults. 

Several of the examined studies build on the initial ideas of Kanewala et al.~\cite{Kanewala2013b}, where they proposed an approach that (a) converts the source code of functions into control-flow graphs, (b) selects source code elements as features for a data set, (c) train a model that can predict whether a feature exhibits a particular metamorphic relation (selected from a pre-compiled list of relations). This requires a set of training data, where features are labeled with a binary classification based on whether or not they exhibit that particular relation. A SVM and Decision Trees are used to train the predictive model. Kanewala et al. extended this work by adding a graph kernel to the process~\cite{Kanewala2016}. Hardin et al. adapted this approach to work with a semi-supervised label propagation algorithm~\cite{Hardin2018}. Finally, Zhang et al.~\cite{Zhang2017} experimented with the use of a RBF NN. They extended the approach to a multi-label classification that can handle multiple metamorphic relations at once instead of predicting one at a time. All four of these studies are evaluated on a variety of functions, from mathematical functions, to data structures, to sorting operations. They were evaluated either using the mutation score or accuracy measurements. 

Nair et al.~\cite{Nair2019} extended this work by demonstrating how data augmentation can enlarge the training dataset by using mutants as the source of the additional training data. They compared the enlarged dataset to the original dataset on a set of 45 matrix calculation functions in terms of the Receiver Operating Characteristic, or the ratio of true positive to false positive classifications.  

Hiremath et al.~\cite{Hiremath2020} propose an approach for using an ML algorithm to predict metamorphic relations for an ocean modeling application. The approach would post a set of relations, evaluate whether they hold, and attempt to minimize a cost function based on the validity of the set of proposed relations. They do not specify an approach, but this maps to common applications of Reinforcement Learning. They do not evaluate their approach, but plan to develop and evaluate it in future work.

We can answer RQ3-5 as follows:

\begin{center}
\begin{framed}
\vspace{-5pt}
  \textbf{RQ3 (Integration of ML):} Almost all studies (96\%) employed a supervised or semi-supervised approach, trained on labeled system execution logs or source code metadata and validated using the accuracy of the trained model.
  \vspace{-5pt}
\end{framed}
\end{center}

\begin{center}
\begin{framed}
\vspace{-5pt}
  \textbf{RQ4 (ML Techniques):} 59\% of the approaches employed a NN---including Backpropagation NNs (27\%), Multilayer Perceptrons (18\%), RBF NN (9\%), probabilistic NN (5\%), and Deep NN (5\%). 23\% of approaches adopted support vector machines. One also adopted decision trees (5\%), and used adaptive boosting (5\%). 5\% did not specify a technique.
  \vspace{-5pt}
\end{framed}
\end{center}

\begin{center}
\begin{framed}
\vspace{-10pt}
  \item \textbf{RQ5 (Evaluation of Approach):} Results were most often evaluated using the mutation score (55\%), followed by number of correct classifications (18\%), classification accuracy (18\%), and ROC (5\%). One study did not perform evaluation.
  \vspace{-5pt}
\end{framed}
\end{center}

\subsection{Limitations and Open Challenges}\label{sec:challenges}

The sampled studies show great promise. They illustrate the potential for solving the oracle problem. However, we have observed multiple limitations and challenges that must be overcome to transition research into use in real-world software development.

\smallskip\noindent\textbf{Volume, Contents, and Collection of Training Data:} Supervised ML approaches, even semi-supervised approaches, require training data to create the predictive model that serves as the test oracle. There are multiple challenges related to the \textit{required volume} of training data, the \textit{required contents} of the training data, and the \textit{human effort} required to produce that training data.

Regardless of the specific type of oracle, the volume of training data that is needed can be vast. This data is generally attained from labeled system execution logs, which means that the SUT needs to be executed \textit{many} times to gather the information needed to train the model. Approaches based on deep learning could produce highly accurate oracles, but may require thousands of executions to gather the required training data. Some approaches also must preprocess the collected data before training. The time required to produce the training data can be high and must be considered. 

This is particularly true for expected value oracles. Even if the output is abstracted into a small pool of representative values, predicting one of several values is a more difficult task than a boolean classification, and requires significant training data for \textit{each} of the values that can result to make accurate classifications. In addition, the training data for expected value oracles must come from passing test cases---i.e., the output must be what was expected---or labels must be hand-applied by humans. A small number of cases based on failing output may be acceptable if the algorithm is resilient to noise in the training data, but training on faulty code can easily result in an inaccurate model. This introduces a significant barrier to automating training by, e.g., generating test input and simply recording the output that results.

Oracles that produce a direct test verdict model a simpler classification problem---is the result a pass or a fail? However, the requirements on the contents of the underlying data are significant. Each entry in the dataset must be assigned a verdict in order to train the model. This requires either existing test oracles---reducing the need for a ML-based oracle in the first place---or human labeling of test results. Humans are limited in their ability to serve as an oracle, as judging test results is time-consuming and can be erroneous as tester becomes fatigued~\cite{McMinn10:humanoracle,Harman13:oraclesurvey}. This makes it difficult to produce a significant volume of training data. Further complicating this problem is the fact that training a test verdict oracle requires the training data to contain a large number of \textit{failing test cases}. This implies that faults have already been discovered in the system and, presumably, fixed before the oracle is trained. This also will reduce the potential effectiveness of a ML-based oracle. 

Metamorphic relation oracles face a similar dilemma. In many of the approaches, the training data consists of source code features labeled with a classification representing whether a particular type of metamorphic relation holds over that feature. This training data must be hand-labeled by a human tester with knowledge of whether these relations hold or not. This requires significant up-front knowledge and effort to establish the ground truth. 

Regardless of the oracle type, generating oracles for complex systems will require ML techniques that can extrapolate from limited training data and that can tolerate noise in the training data. Means of generating synthetic training data, like in the work of Nair et al.~\cite{Nair2019}, demonstrate the potential for \textit{data augmentation} to help in overcoming this limitation. 

\smallskip\noindent\textbf{Retraining and Feedback:} After training, models generated by supervised learning techniques have a fixed error rate and do not learn from new mistakes made after training. In other words, if the training data is insufficient or inaccurate, the generated oracle will remain inaccurate as long as it remains in use. The ability to improve the oracle based on additional feedback after training could help account for limitations in the initial training data. 

There are two primary means to overcome this limitation---either retraining the model using an enriched training dataset, or adopting a reinforcement learning approach that can adapt its expectations based on attained feedback on the accuracy of its decisions. Both means carry challenges. Retraining requires (a) establishing a schedule for when to train the updated model, and (b), an active effort on the part of human testers to enrich and curate the training dataset. Enriching this dataset---as well as the use of RL---requires some kind of feedback mechanism to judge the accuracy of the oracle. This is likely to require human feedback on, at least, a subset of the decisions made, reducing the potential cost savings. 

\begin{center}
\begin{framed}
\vspace{-10pt}
  \item \textbf{RQ6 (Challenges):} Oracle generation is limited by the required quantity, quality, and content of training data. Assembling training data may require significant human effort. Models should be retrained over time.
  \vspace{-5pt}
\end{framed}
\end{center}

\smallskip\noindent\textbf{Complexity of Modeled Functionality:} Many approaches are  \newline demonstrated on highly simplistic functions, with only a few lines of code and a small number of possible outputs. While it is intuitive to \textit{start} with highly simplistic examples to examine the viability of an approach, application of such techniques in the field would require oracle generation for far more complex system functions. If a function is simple, there is likely little need for oracle generation in the first place. It remains to be seen whether generated oracles can predict the output of real-world production code, or even simple code with an unconstrained or lightly constrained output space. 

Generation of an expected output oracle that can model any arbitrary function with unconstrained output may be prohibitively difficulty for even the most effective ML techniques available today. Some abstraction should be expected. One possibility to consider is a variable level of abstraction---e.g., a training-time decision to cluster the output into an adjustable number of representative values (i.e., the centroid of each cluster). Training could take place over different settings for this parameter, and an acceptable balance between quality and level-of-detail could be explored.   

\smallskip\noindent\textbf{Variety, Complexity, and Tuning of ML Techniques:} Many of the proposed approaches---especially the earlier ones---are based on simple neural networks with only a few hidden layers. These techniques have strict limitations in the complexity of the functions they can model, and have been superseded by newer ML techniques. Deep learning techniques, which may utilize a high number of hidden layers, may be key in building models of more complex functions. One approach to date has utilized deep learning~\cite{Monsefi2019}, and we would expect more to explore these techniques in the coming years. However, deep learning also introduces steep requirements on the training data that may limit its applicability~\cite{taylor2017improving}. 

Almost all of the proposed approaches are based on a single ML technique. An approach explored in other domains is the use of \textit{ensembles}~\cite{Kocaguneli12:Ensemble}. In such approaches, models are trained on the same data using a variety of techniques. Each model is asked for a prediction, then the final prediction is based on the consensus of the ensemble. Ensembles are often able to reach stable, accurate conclusions in situations where a single model may be inaccurate. Ensembles may be a way to overcome the fragility of current oracle generation approaches. 

Many ML techniques have a number of \textit{hyperparameters} that can be tuned (e.g., the learning rate, number of hidden units, or activation function)~\cite{Minku19:Hyperparameter}. Hyperparameter tuning can have a major impact on model accuracy, and can enable significant improvements in the results of even simple ML techniques. The proposed approaches do not explore the impact of hyperparameter tuning on the trained models. This is an oversight that should be corrected in future work.

\begin{center}
\begin{framed}
\vspace{-10pt}
  \item \textbf{RQ6 (Challenges):} Applied techniques may be insufficient for modeling complex functions with many possible outputs. Varying degrees of output abstraction should be explored. Deep learning and ensemble techniques, as well as hyperparameter tuning, should be explored.
  \vspace{-5pt}
\end{framed}
\end{center}

\smallskip\noindent\textbf{Lack of a Standard Benchmark:} The emergence of bug benchmarks (e.g.,~\cite{Gay20:D4J,Saha18:BugsJar}) has enabled sophisticated analyses and comparison of approaches to automated input generation and program repair. To date, oracle generation has often been evaluated on case examples---often over-simplistic examples---where code or metadata is unavailable. This makes comparison and replication difficult. 

The creation of a benchmark for oracle generation research could advance the state-of-the-art in the field, spur new research advances, and enable replication and extension of proposed approaches. Such a benchmark should contain a variety of code examples from multiple domains and of varying levels of complexity, allowing the field to move beyond over-simplistic examples. Code examples should be paired with the metadata needed to support oracle generation. This would include sample test cases and human-created test oracles, at minimum. Such a benchmark could also include sample training data that could be augmented over time by researchers. 

\smallskip\noindent\textbf{Lack of Replication Package or Open Code:} A common dilemma in software engineering research is lack of access to the code built by researchers or the data used to draw conclusions. Often, the paper itself is not sufficient to allow replication or application of the technique in a new context. This applies to  research in oracle generation as well. Some studies make use of open-source ML frameworks (e.g., scikit-learn). This is positive, in that the tools are trustworthy and available. However, without the authors' code and data, there may not be enough information to enable replication. Further, these frameworks themselves evolve over time, and the attained results may differ because the underlying ML technique has changed since the original study was published. 

New approaches should include a replication package with the source code written by the authors, execution scripts, and the versions of external dependencies that were used at the time that the study was performed. This should also include data used by the authors in their analyses. 

\begin{center}
\begin{framed}
\vspace{-10pt}
  \item \textbf{RQ6 (Challenges):} Research is limited by overuse of simplistic examples, the lack of common benchmarks, and the unavailability of code and data. A benchmark should be created for evaluating oracle research, and researchers should be encouraged to provide replication packages and open code.
  \vspace{-5pt}
\end{framed}
\end{center}

\section{Threats to Validity}\label{sec:threats}

\noindent\textbf{External and Internal Validity:} Our conclusions are based on the studies sampled. It is possible that we may have omitted important studies or sampled an inadequate number of studies. This can affect internal validity---the evidence we use to make conclusions---and external validity---the generalizability of our findings. SLRs are not required to reflect all studies from a research field. Rather, their selection protocol (search string, inclusion and exclusion criteria) should be sufficient to ensure an adequate sample of the field. We believe that our selection strategy was appropriate. We tested different search strings, and performed a validation exercise to test the robustness of our string. We have used four databases, covering the majority of relevant software engineering venues. Our final set of studies includes 22 primary studies, which we believe is sufficient to make informed conclusions. 

\smallskip\noindent\textbf{Conclusion Validity:} The analyses performed are qualitative, and require inference from the authors. This could introduce bias into our conclusions. For example, subjective judgements are required as part of article selection, data extraction, and coding (e.g., categorizing studies based on the oracle type). To control for bias, protocols were discussed and agreed upon by both authors, and independent verification took place on---at least---a sample of all decisions made by either author. 

\smallskip\noindent\textbf{Construct Validity:} We used a set of properties to guide data extraction. These properties may have been incomplete or misleading. However, we have tried to establish properties that were appropriate and directly informed by our research questions. These properties were iteratively refined using a selection of papers. 

\section{Conclusions} \label{sec:conclusions}

Machine learning has the potential to solve the \textit{``test oracle problem''}---the challenge of automatically generating oracles for a function. We have characterized emerging research in this area through a systematic literature review examining oracle types, researcher goals, the ML techniques applied, how the generation process was assessed, and the open research challenges in this emerging field.

Based on 22 relevant studies, we observed that ML algorithms have been used to generate test verdict, metamorphic relation, and---most commonly---expected output oracles. The ML algorithms train predictive models that serve either as a stand-in for an existing test oracle---predicting a test verdict---or as a way to learn information about a function---either the expected output or metamorphic relations---that can be used as part of issuing a verdict.

Almost all studies employed a supervised or semi-supervised approach, trained on labeled system executions or source code metadata. Of these approaches, many used some type of neural network---including Backpropagation NNs, Multilayer Perceptrons, RBF NNs, probabilistic NNs, and Deep NNs. Others applied include support vector machines, decision trees, and adaptive boosting. Results were most often evaluated using the mutation score, number of correct classifications, classification accuracy, and ROC. 

The studies show great promise, but there are significant open challenges. Generation is limited by the required quantity, quality, and content of training data. Models should be retrained over time. Applied techniques may be insufficient for modeling complex functions with many possible outputs. Varying degrees of output abstraction, deep learning and ensemble techniques, and hyperparameter tuning should be explored. In addition, research is limited by overuse of simplistic examples, lack of common benchmarks, and unavailability of code and data. A robust open benchmark should be created, and researchers should provide replication packages. 

\begin{acks}
This research was supported by Vetenskapsr{\aa}det grant 2019-05275. 
\end{acks}

\bibliographystyle{ACM-Reference-Format}
\bibliography{refs}


\begin{thebibliography}{41}


\ifx \showCODEN    \undefined \def \showCODEN     #1{\unskip}     \fi
\ifx \showDOI      \undefined \def \showDOI       #1{#1}\fi
\ifx \showISBNx    \undefined \def \showISBNx     #1{\unskip}     \fi
\ifx \showISBNxiii \undefined \def \showISBNxiii  #1{\unskip}     \fi
\ifx \showISSN     \undefined \def \showISSN      #1{\unskip}     \fi
\ifx \showLCCN     \undefined \def \showLCCN      #1{\unskip}     \fi
\ifx \shownote     \undefined \def \shownote      #1{#1}          \fi
\ifx \showarticletitle \undefined \def \showarticletitle #1{#1}   \fi
\ifx \showURL      \undefined \def \showURL       {\relax}        \fi
\providecommand\bibfield[2]{#2}
\providecommand\bibinfo[2]{#2}
\providecommand\natexlab[1]{#1}
\providecommand\showeprint[2][]{arXiv:#2}

\bibitem[\protect\citeauthoryear{Aggarwal, Singh, Kaur, and Sangwan}{Aggarwal
  et~al\mbox{.}}{2004}]%
        {Aggarwal2004}
\bibfield{author}{\bibinfo{person}{K.~K. Aggarwal}, \bibinfo{person}{Yogesh
  Singh}, \bibinfo{person}{A. Kaur}, {and} \bibinfo{person}{O.~P. Sangwan}.}
  \bibinfo{year}{2004}\natexlab{}.
\newblock \showarticletitle{A Neural Net Based Approach to Test Oracle}.
\newblock \bibinfo{journal}{\emph{SIGSOFT Softw. Eng. Notes}}
  \bibinfo{volume}{29}, \bibinfo{number}{3} (\bibinfo{date}{May}
  \bibinfo{year}{2004}), \bibinfo{pages}{1–6}.
\newblock
\showISSN{0163-5948}
\urldef\tempurl%
\url{https://doi.org/10.1145/986710.986725}
\showDOI{\tempurl}


\bibitem[\protect\citeauthoryear{Almasi, Hemmati, Fraser, Arcuri, and
  Benefelds}{Almasi et~al\mbox{.}}{2017}]%
        {Almasi17:IndustrialEval}
\bibfield{author}{\bibinfo{person}{M.~Moein Almasi}, \bibinfo{person}{Hadi
  Hemmati}, \bibinfo{person}{Gordon Fraser}, \bibinfo{person}{Andrea Arcuri},
  {and} \bibinfo{person}{Janis Benefelds}.} \bibinfo{year}{2017}\natexlab{}.
\newblock \showarticletitle{An Industrial Evaluation of Unit Test Generation:
  Finding Real Faults in a Financial Application}. In
  \bibinfo{booktitle}{\emph{Proceedings of the 39th IEEE/ACM International
  Conference on Software Engineering (ICSE)---Software Engineering in Practice
  Track (SEIP)}} (Buenos Aires, Argentina) \emph{(\bibinfo{series}{ICSE
  2017})}. \bibinfo{publisher}{ACM}, \bibinfo{address}{New York, NY, USA}.
\newblock


\bibitem[\protect\citeauthoryear{Alpaydin}{Alpaydin}{2020}]%
        {MLbook2020}
\bibfield{author}{\bibinfo{person}{Ethem Alpaydin}.}
  \bibinfo{year}{2020}\natexlab{}.
\newblock \bibinfo{booktitle}{\emph{Introduction to machine learning}}.
\newblock \bibinfo{publisher}{MIT press}.
\newblock


\bibitem[\protect\citeauthoryear{Barr, Harman, McMinn, Shahbaz, and Yoo}{Barr
  et~al\mbox{.}}{2015}]%
        {Harman13:oraclesurvey}
\bibfield{author}{\bibinfo{person}{E.T. Barr}, \bibinfo{person}{M. Harman},
  \bibinfo{person}{P. McMinn}, \bibinfo{person}{M. Shahbaz}, {and}
  \bibinfo{person}{Shin Yoo}.} \bibinfo{year}{2015}\natexlab{}.
\newblock \showarticletitle{The Oracle Problem in Software Testing: A Survey}.
\newblock \bibinfo{journal}{\emph{IEEE Transactions on Software Engineering}}
  \bibinfo{volume}{41}, \bibinfo{number}{5} (\bibinfo{date}{May}
  \bibinfo{year}{2015}), \bibinfo{pages}{507--525}.
\newblock
\showISSN{0098-5589}
\urldef\tempurl%
\url{https://doi.org/10.1109/TSE.2014.2372785}
\showDOI{\tempurl}


\bibitem[\protect\citeauthoryear{Braga, Neto, Rab\^{e}lo, Santiago, and
  Souza}{Braga et~al\mbox{.}}{2018}]%
        {Braga2018}
\bibfield{author}{\bibinfo{person}{Rony\'{e}rison Braga},
  \bibinfo{person}{Pedro~Santos Neto}, \bibinfo{person}{Ricardo Rab\^{e}lo},
  \bibinfo{person}{Jos\'{e} Santiago}, {and} \bibinfo{person}{Matheus Souza}.}
  \bibinfo{year}{2018}\natexlab{}.
\newblock \showarticletitle{A Machine Learning Approach to Generate Test
  Oracles}. In \bibinfo{booktitle}{\emph{Proceedings of the XXXII Brazilian
  Symposium on Software Engineering}} (Sao Carlos, Brazil)
  \emph{(\bibinfo{series}{SBES '18})}. \bibinfo{publisher}{Association for
  Computing Machinery}, \bibinfo{address}{New York, NY, USA},
  \bibinfo{pages}{142–151}.
\newblock
\showISBNx{9781450365031}
\urldef\tempurl%
\url{https://doi.org/10.1145/3266237.3266273}
\showDOI{\tempurl}


\bibitem[\protect\citeauthoryear{Ding and Zhang}{Ding and Zhang}{2016}]%
        {Ding2016}
\bibfield{author}{\bibinfo{person}{J. Ding} {and} \bibinfo{person}{D. Zhang}.}
  \bibinfo{year}{2016}\natexlab{}.
\newblock \showarticletitle{A machine learning approach for developing test
  oracles for testing scientific software}.
\newblock \bibinfo{journal}{\emph{Proceedings of the International Conference
  on Software Engineering and Knowledge Engineering, SEKE}}
  \bibinfo{volume}{2016-January} (\bibinfo{year}{2016}),
  \bibinfo{pages}{390--395}.
\newblock
\urldef\tempurl%
\url{https://doi.org/10.18293/SEKE2016-137}
\showDOI{\tempurl}
\newblock
\shownote{cited By 4.}


\bibitem[\protect\citeauthoryear{Dunjko and Briegel}{Dunjko and
  Briegel}{2018}]%
        {MLintro1}
\bibfield{author}{\bibinfo{person}{Vedran Dunjko} {and} \bibinfo{person}{Hans~J
  Briegel}.} \bibinfo{year}{2018}\natexlab{}.
\newblock \showarticletitle{Machine learning \& artificial intelligence in the
  quantum domain: a review of recent progress}.
\newblock \bibinfo{journal}{\emph{Reports on Progress in Physics}}
  \bibinfo{volume}{81}, \bibinfo{number}{7} (\bibinfo{year}{2018}),
  \bibinfo{pages}{074001}.
\newblock


\bibitem[\protect\citeauthoryear{Durelli, Durelli, Borges, Endo, Eler, Dias,
  and Guimaraes}{Durelli et~al\mbox{.}}{2019}]%
        {surveyMLinTesting2019}
\bibfield{author}{\bibinfo{person}{Vinicius~HS Durelli},
  \bibinfo{person}{Rafael~S Durelli}, \bibinfo{person}{Simone~S Borges},
  \bibinfo{person}{Andre~T Endo}, \bibinfo{person}{Marcelo~M Eler},
  \bibinfo{person}{Diego~RC Dias}, {and} \bibinfo{person}{Marcelo~P
  Guimaraes}.} \bibinfo{year}{2019}\natexlab{}.
\newblock \showarticletitle{Machine learning applied to software testing: A
  systematic mapping study}.
\newblock \bibinfo{journal}{\emph{IEEE Transactions on Reliability}}
  \bibinfo{volume}{68}, \bibinfo{number}{3} (\bibinfo{year}{2019}),
  \bibinfo{pages}{1189--1212}.
\newblock


\bibitem[\protect\citeauthoryear{Gay and Just}{Gay and Just}{2020}]%
        {Gay20:D4J}
\bibfield{author}{\bibinfo{person}{Gregory Gay} {and} \bibinfo{person}{Ren{\'e}
  Just}.} \bibinfo{year}{2020}\natexlab{}.
\newblock \showarticletitle{Defects4J as a Challenge Case for the Search-Based
  Software Engineering Community}. In \bibinfo{booktitle}{\emph{Search-Based
  Software Engineering}}, \bibfield{editor}{\bibinfo{person}{Aldeida Aleti}
  {and} \bibinfo{person}{Annibale Panichella}} (Eds.).
  \bibinfo{publisher}{Springer International Publishing},
  \bibinfo{address}{Cham}, \bibinfo{pages}{255--261}.
\newblock
\showISBNx{978-3-030-59762-7}


\bibitem[\protect\citeauthoryear{Gay, Staats, Whalen, and Heimdahl}{Gay
  et~al\mbox{.}}{2015a}]%
        {Gay15:oracleselection}
\bibfield{author}{\bibinfo{person}{G. Gay}, \bibinfo{person}{M. Staats},
  \bibinfo{person}{M. Whalen}, {and} \bibinfo{person}{M. Heimdahl}.}
  \bibinfo{year}{2015}\natexlab{a}.
\newblock \showarticletitle{Automated Oracle Data Selection Support}.
\newblock \bibinfo{journal}{\emph{Software Engineering, IEEE Transactions on}}
  \bibinfo{volume}{PP}, \bibinfo{number}{99} (\bibinfo{year}{2015}),
  \bibinfo{pages}{1--1}.
\newblock
\showISSN{0098-5589}
\urldef\tempurl%
\url{https://doi.org/10.1109/TSE.2015.2436920}
\showDOI{\tempurl}


\bibitem[\protect\citeauthoryear{Gay, Staats, Whalen, and Heimdahl}{Gay
  et~al\mbox{.}}{2015b}]%
        {Gay15:risks}
\bibfield{author}{\bibinfo{person}{G. Gay}, \bibinfo{person}{M. Staats},
  \bibinfo{person}{M. Whalen}, {and} \bibinfo{person}{M.P.E. Heimdahl}.}
  \bibinfo{year}{2015}\natexlab{b}.
\newblock \showarticletitle{The Risks of Coverage-Directed Test Case
  Generation}.
\newblock \bibinfo{journal}{\emph{Software Engineering, IEEE Transactions on}}
  \bibinfo{volume}{PP}, \bibinfo{number}{99} (\bibinfo{year}{2015}).
\newblock
\showISSN{0098-5589}
\urldef\tempurl%
\url{https://doi.org/10.1109/TSE.2015.2421011}
\showDOI{\tempurl}


\bibitem[\protect\citeauthoryear{{Gholami}, {Attar}, {Haghighi}, {Asl},
  {Valueian}, and {Mohamadyari}}{{Gholami} et~al\mbox{.}}{2018}]%
        {Gholami2018}
\bibfield{author}{\bibinfo{person}{F. {Gholami}}, \bibinfo{person}{N. {Attar}},
  \bibinfo{person}{H. {Haghighi}}, \bibinfo{person}{M.~V. {Asl}},
  \bibinfo{person}{M. {Valueian}}, {and} \bibinfo{person}{S. {Mohamadyari}}.}
  \bibinfo{year}{2018}\natexlab{}.
\newblock \showarticletitle{A classifier-based test oracle for embedded
  software}. In \bibinfo{booktitle}{\emph{2018 Real-Time and Embedded Systems
  and Technologies (RTEST)}}. \bibinfo{pages}{104--111}.
\newblock
\urldef\tempurl%
\url{https://doi.org/10.1109/RTEST.2018.8397165}
\showDOI{\tempurl}


\bibitem[\protect\citeauthoryear{Goodfellow, Bengio, Courville, and
  Bengio}{Goodfellow et~al\mbox{.}}{2016}]%
        {DLbook2016}
\bibfield{author}{\bibinfo{person}{Ian Goodfellow}, \bibinfo{person}{Yoshua
  Bengio}, \bibinfo{person}{Aaron Courville}, {and} \bibinfo{person}{Yoshua
  Bengio}.} \bibinfo{year}{2016}\natexlab{}.
\newblock \bibinfo{booktitle}{\emph{Deep learning}}. Vol.~\bibinfo{volume}{1}.
\newblock \bibinfo{publisher}{MIT press Cambridge}.
\newblock


\bibitem[\protect\citeauthoryear{Hardin and Kanewala}{Hardin and
  Kanewala}{2018}]%
        {Hardin2018}
\bibfield{author}{\bibinfo{person}{Bonnie Hardin} {and} \bibinfo{person}{Upulee
  Kanewala}.} \bibinfo{year}{2018}\natexlab{}.
\newblock \showarticletitle{Using Semi-Supervised Learning for Predicting
  Metamorphic Relations}. In \bibinfo{booktitle}{\emph{Proceedings of the 3rd
  International Workshop on Metamorphic Testing}} (Gothenburg, Sweden)
  \emph{(\bibinfo{series}{MET '18})}. \bibinfo{publisher}{Association for
  Computing Machinery}, \bibinfo{address}{New York, NY, USA},
  \bibinfo{pages}{14–17}.
\newblock
\showISBNx{9781450357296}
\urldef\tempurl%
\url{https://doi.org/10.1145/3193977.3193985}
\showDOI{\tempurl}


\bibitem[\protect\citeauthoryear{{Hiremath}, {Claus}, {Hasselbring}, and
  {Rath}}{{Hiremath} et~al\mbox{.}}{2020}]%
        {Hiremath2020}
\bibfield{author}{\bibinfo{person}{D.~J. {Hiremath}}, \bibinfo{person}{M.
  {Claus}}, \bibinfo{person}{W. {Hasselbring}}, {and} \bibinfo{person}{W.
  {Rath}}.} \bibinfo{year}{2020}\natexlab{}.
\newblock \showarticletitle{Automated identification of metamorphic test
  scenarios for an ocean-modeling application}. In
  \bibinfo{booktitle}{\emph{2020 IEEE International Conference On Artificial
  Intelligence Testing (AITest)}}. \bibinfo{pages}{62--63}.
\newblock
\urldef\tempurl%
\url{https://doi.org/10.1109/AITEST49225.2020.00016}
\showDOI{\tempurl}


\bibitem[\protect\citeauthoryear{{Jin}, {Wang}, {Chen}, {Gou}, and
  {Wang}}{{Jin} et~al\mbox{.}}{2008}]%
        {Jin2008}
\bibfield{author}{\bibinfo{person}{H. {Jin}}, \bibinfo{person}{Y. {Wang}},
  \bibinfo{person}{N. {Chen}}, \bibinfo{person}{Z. {Gou}}, {and}
  \bibinfo{person}{S. {Wang}}.} \bibinfo{year}{2008}\natexlab{}.
\newblock \showarticletitle{Artificial Neural Network for Automatic Test
  Oracles Generation}. In \bibinfo{booktitle}{\emph{2008 International
  Conference on Computer Science and Software Engineering}},
  Vol.~\bibinfo{volume}{2}. \bibinfo{pages}{727--730}.
\newblock
\urldef\tempurl%
\url{https://doi.org/10.1109/CSSE.2008.774}
\showDOI{\tempurl}


\bibitem[\protect\citeauthoryear{Kanewala and Bieman}{Kanewala and
  Bieman}{2013}]%
        {Kanewala2013b}
\bibfield{author}{\bibinfo{person}{U. Kanewala} {and} \bibinfo{person}{J.M.
  Bieman}.} \bibinfo{year}{2013}\natexlab{}.
\newblock \showarticletitle{Using machine learning techniques to detect
  metamorphic relations for programs without test oracles}.
\newblock \bibinfo{journal}{\emph{2013 IEEE 24th International Symposium on
  Software Reliability Engineering, ISSRE 2013}} (\bibinfo{year}{2013}),
  \bibinfo{pages}{1--10}.
\newblock
\urldef\tempurl%
\url{https://doi.org/10.1109/ISSRE.2013.6698899}
\showDOI{\tempurl}
\newblock
\shownote{cited By 30.}


\bibitem[\protect\citeauthoryear{Kanewala, Bieman, and Ben-Hur}{Kanewala
  et~al\mbox{.}}{2016}]%
        {Kanewala2016}
\bibfield{author}{\bibinfo{person}{U. Kanewala}, \bibinfo{person}{J.M. Bieman},
  {and} \bibinfo{person}{A. Ben-Hur}.} \bibinfo{year}{2016}\natexlab{}.
\newblock \showarticletitle{Predicting metamorphic relations for testing
  scientific software: A machine learning approach using graph kernels}.
\newblock \bibinfo{journal}{\emph{Software Testing Verification and
  Reliability}} \bibinfo{volume}{26}, \bibinfo{number}{3}
  (\bibinfo{year}{2016}), \bibinfo{pages}{245--269}.
\newblock
\urldef\tempurl%
\url{https://doi.org/10.1002/stvr.1594}
\showDOI{\tempurl}
\newblock
\shownote{cited By 35.}


\bibitem[\protect\citeauthoryear{Kitchenham and Charters}{Kitchenham and
  Charters}{2007}]%
        {Kitchenham07:Guidelines}
\bibfield{author}{\bibinfo{person}{B. Kitchenham} {and} \bibinfo{person}{S
  Charters}.} \bibinfo{year}{2007}\natexlab{}.
\newblock \bibinfo{title}{Guidelines for performing Systematic Literature
  Reviews in Software Engineering}.
\newblock
\newblock


\bibitem[\protect\citeauthoryear{Kocaguneli, Menzies, and Keung}{Kocaguneli
  et~al\mbox{.}}{2012}]%
        {Kocaguneli12:Ensemble}
\bibfield{author}{\bibinfo{person}{Ekrem Kocaguneli}, \bibinfo{person}{Tim
  Menzies}, {and} \bibinfo{person}{Jacky~W. Keung}.}
  \bibinfo{year}{2012}\natexlab{}.
\newblock \showarticletitle{On the Value of Ensemble Effort Estimation}.
\newblock \bibinfo{journal}{\emph{IEEE Transactions on Software Engineering}}
  \bibinfo{volume}{38}, \bibinfo{number}{6} (\bibinfo{year}{2012}),
  \bibinfo{pages}{1403--1416}.
\newblock
\urldef\tempurl%
\url{https://doi.org/10.1109/TSE.2011.111}
\showDOI{\tempurl}


\bibitem[\protect\citeauthoryear{{Makondo}, {Nallanthighal}, {Mapanga}, and
  {Kadebu}}{{Makondo} et~al\mbox{.}}{2016}]%
        {Makondo2016}
\bibfield{author}{\bibinfo{person}{W. {Makondo}}, \bibinfo{person}{R.
  {Nallanthighal}}, \bibinfo{person}{I. {Mapanga}}, {and} \bibinfo{person}{P.
  {Kadebu}}.} \bibinfo{year}{2016}\natexlab{}.
\newblock \showarticletitle{Exploratory test oracle using multi-layer
  perceptron neural network}. In \bibinfo{booktitle}{\emph{2016 International
  Conference on Advances in Computing, Communications and Informatics
  (ICACCI)}}. \bibinfo{pages}{1166--1171}.
\newblock
\urldef\tempurl%
\url{https://doi.org/10.1109/ICACCI.2016.7732202}
\showDOI{\tempurl}


\bibitem[\protect\citeauthoryear{McMinn, Stevenson, and Harman}{McMinn
  et~al\mbox{.}}{2010}]%
        {McMinn10:humanoracle}
\bibfield{author}{\bibinfo{person}{Phil McMinn}, \bibinfo{person}{Mark
  Stevenson}, {and} \bibinfo{person}{Mark Harman}.}
  \bibinfo{year}{2010}\natexlab{}.
\newblock \showarticletitle{Reducing Qualitative Human Oracle Costs Associated
  with Automatically Generated Test Data}. In
  \bibinfo{booktitle}{\emph{Proceedings of the First International Workshop on
  Software Test Output Validation}} (Trento, Italy)
  \emph{(\bibinfo{series}{STOV '10})}. \bibinfo{publisher}{ACM},
  \bibinfo{address}{New York, NY, USA}, \bibinfo{pages}{1--4}.
\newblock
\showISBNx{978-1-4503-0138-1}
\urldef\tempurl%
\url{https://doi.org/10.1145/1868048.1868049}
\showDOI{\tempurl}


\bibitem[\protect\citeauthoryear{Minku}{Minku}{2019}]%
        {Minku19:Hyperparameter}
\bibfield{author}{\bibinfo{person}{Leandro~L. Minku}.}
  \bibinfo{year}{2019}\natexlab{}.
\newblock \showarticletitle{A novel online supervised hyperparameter tuning
  procedure applied to cross-company software effort estimation}.
\newblock \bibinfo{journal}{\emph{Empirical Software Engineering}}
  \bibinfo{volume}{24}, \bibinfo{number}{5} (\bibinfo{year}{2019}),
  \bibinfo{pages}{3153--3204}.
\newblock
\showISBNx{1573-7616}
\urldef\tempurl%
\url{https://doi.org/10.1007/s10664-019-09686-w}
\showDOI{\tempurl}


\bibitem[\protect\citeauthoryear{Monsefi, Zakeri, Samsam, and
  Khashehchi}{Monsefi et~al\mbox{.}}{2019}]%
        {Monsefi2019}
\bibfield{author}{\bibinfo{person}{A.K. Monsefi}, \bibinfo{person}{B. Zakeri},
  \bibinfo{person}{S. Samsam}, {and} \bibinfo{person}{M. Khashehchi}.}
  \bibinfo{year}{2019}\natexlab{}.
\newblock \showarticletitle{Performing Software Test Oracle Based on Deep
  Neural Network with Fuzzy Inference System}.
\newblock \bibinfo{journal}{\emph{Communications in Computer and Information
  Science}}  \bibinfo{volume}{891} (\bibinfo{year}{2019}),
  \bibinfo{pages}{406--417}.
\newblock
\urldef\tempurl%
\url{https://doi.org/10.1007/978-3-030-33495-6_31}
\showDOI{\tempurl}
\newblock
\shownote{cited By 0.}


\bibitem[\protect\citeauthoryear{Naik and Tripathy}{Naik and Tripathy}{2011}]%
        {STintro2011}
\bibfield{author}{\bibinfo{person}{Kshirasagar Naik} {and}
  \bibinfo{person}{Priyadarshi Tripathy}.} \bibinfo{year}{2011}\natexlab{}.
\newblock \bibinfo{booktitle}{\emph{Software testing and quality assurance:
  theory and practice}}.
\newblock \bibinfo{publisher}{John Wiley \& Sons}.
\newblock


\bibitem[\protect\citeauthoryear{Nair, Meinke, and Eldh}{Nair
  et~al\mbox{.}}{2019}]%
        {Nair2019}
\bibfield{author}{\bibinfo{person}{Aravind Nair}, \bibinfo{person}{Karl
  Meinke}, {and} \bibinfo{person}{Sigrid Eldh}.}
  \bibinfo{year}{2019}\natexlab{}.
\newblock \showarticletitle{Leveraging Mutants for Automatic Prediction of
  Metamorphic Relations Using Machine Learning}. In
  \bibinfo{booktitle}{\emph{Proceedings of the 3rd ACM SIGSOFT International
  Workshop on Machine Learning Techniques for Software Quality Evaluation}}
  (Tallinn, Estonia) \emph{(\bibinfo{series}{MaLTeSQuE 2019})}.
  \bibinfo{publisher}{Association for Computing Machinery},
  \bibinfo{address}{New York, NY, USA}, \bibinfo{pages}{1–6}.
\newblock
\showISBNx{9781450368551}
\urldef\tempurl%
\url{https://doi.org/10.1145/3340482.3342741}
\showDOI{\tempurl}


\bibitem[\protect\citeauthoryear{Orso and Rothermel}{Orso and
  Rothermel}{2014}]%
        {Orso14:STR}
\bibfield{author}{\bibinfo{person}{Alessandro Orso} {and}
  \bibinfo{person}{Gregg Rothermel}.} \bibinfo{year}{2014}\natexlab{}.
\newblock \showarticletitle{Software Testing: A Research Travelogue
  (2000--2014)}. In \bibinfo{booktitle}{\emph{Proceedings of the on Future of
  Software Engineering}} (Hyderabad, India) \emph{(\bibinfo{series}{FOSE
  2014})}. \bibinfo{publisher}{ACM}, \bibinfo{address}{New York, NY, USA},
  \bibinfo{pages}{117--132}.
\newblock
\showISBNx{978-1-4503-2865-4}
\urldef\tempurl%
\url{https://doi.org/10.1145/2593882.2593885}
\showDOI{\tempurl}


\bibitem[\protect\citeauthoryear{{Richard S. Sutton and Andrew G.
  Barto}}{{Richard S. Sutton and Andrew G. Barto}}{2018}]%
        {Sutton2018}
\bibfield{author}{\bibinfo{person}{{Richard S. Sutton and Andrew G. Barto}}.}
  \bibinfo{year}{2018}\natexlab{}.
\newblock \bibinfo{booktitle}{\emph{{Reinforcement Learning, Second Edition An
  Introduction}}}.
\newblock 550 pages.
\newblock
\showISBNx{9780262039246}


\bibitem[\protect\citeauthoryear{Richardson, Aha, and O'Malley}{Richardson
  et~al\mbox{.}}{1992}]%
        {Richardson92:TestOracles}
\bibfield{author}{\bibinfo{person}{D.~J. Richardson}, \bibinfo{person}{S.~L.
  Aha}, {and} \bibinfo{person}{T. O'Malley}.} \bibinfo{year}{1992}\natexlab{}.
\newblock \showarticletitle{Specification-based Test Oracles for Reactive
  Systems}. In \bibinfo{booktitle}{\emph{Proc. of the 14th Int'l Conf. on
  Software Engineering}}. \bibinfo{publisher}{Springer},
  \bibinfo{pages}{105--118}.
\newblock


\bibitem[\protect\citeauthoryear{Saha, Lyu, Lam, Yoshida, and Prasad}{Saha
  et~al\mbox{.}}{2018}]%
        {Saha18:BugsJar}
\bibfield{author}{\bibinfo{person}{Ripon~K. Saha}, \bibinfo{person}{Yingjun
  Lyu}, \bibinfo{person}{Wing Lam}, \bibinfo{person}{Hiroaki Yoshida}, {and}
  \bibinfo{person}{Mukul~R. Prasad}.} \bibinfo{year}{2018}\natexlab{}.
\newblock \showarticletitle{Bugs.Jar: A Large-Scale, Diverse Dataset of
  Real-World Java Bugs}. In \bibinfo{booktitle}{\emph{Proceedings of the 15th
  International Conference on Mining Software Repositories}} (Gothenburg,
  Sweden) \emph{(\bibinfo{series}{MSR '18})}. \bibinfo{publisher}{Association
  for Computing Machinery}, \bibinfo{address}{New York, NY, USA},
  \bibinfo{pages}{10–13}.
\newblock
\showISBNx{9781450357166}
\urldef\tempurl%
\url{https://doi.org/10.1145/3196398.3196473}
\showDOI{\tempurl}


\bibitem[\protect\citeauthoryear{Sangwan, Bhatia, and Singh}{Sangwan
  et~al\mbox{.}}{2011}]%
        {Sangwan2011}
\bibfield{author}{\bibinfo{person}{Om~Prakash Sangwan},
  \bibinfo{person}{Pradeep~Kumar Bhatia}, {and} \bibinfo{person}{Yogesh
  Singh}.} \bibinfo{year}{2011}\natexlab{}.
\newblock \showarticletitle{Radial Basis Function Neural Network Based Approach
  to Test Oracle}.
\newblock \bibinfo{journal}{\emph{SIGSOFT Softw. Eng. Notes}}
  \bibinfo{volume}{36}, \bibinfo{number}{5} (\bibinfo{date}{Sept.}
  \bibinfo{year}{2011}), \bibinfo{pages}{1–5}.
\newblock
\showISSN{0163-5948}
\urldef\tempurl%
\url{https://doi.org/10.1145/2020976.2020992}
\showDOI{\tempurl}


\bibitem[\protect\citeauthoryear{Shahamiri, Kadir, Ibrahim, and
  Hashim}{Shahamiri et~al\mbox{.}}{2011}]%
        {Shahamiri2011}
\bibfield{author}{\bibinfo{person}{S.R. Shahamiri}, \bibinfo{person}{W.M.N.W.
  Kadir}, \bibinfo{person}{S. Ibrahim}, {and} \bibinfo{person}{S.Z.M. Hashim}.}
  \bibinfo{year}{2011}\natexlab{}.
\newblock \showarticletitle{An automated framework for software test oracle}.
\newblock \bibinfo{journal}{\emph{Information and Software Technology}}
  \bibinfo{volume}{53}, \bibinfo{number}{7} (\bibinfo{year}{2011}),
  \bibinfo{pages}{774--788}.
\newblock
\urldef\tempurl%
\url{https://doi.org/10.1016/j.infsof.2011.02.006}
\showDOI{\tempurl}
\newblock
\shownote{cited By 31.}


\bibitem[\protect\citeauthoryear{Shahamiri, Wan~Kadir, and
  Bin~Ibrahim}{Shahamiri et~al\mbox{.}}{2010}]%
        {Shahamiri2010a}
\bibfield{author}{\bibinfo{person}{S.R. Shahamiri}, \bibinfo{person}{W.M.N.
  Wan~Kadir}, {and} \bibinfo{person}{S. Bin~Ibrahim}.}
  \bibinfo{year}{2010}\natexlab{}.
\newblock \showarticletitle{An automated oracle approach to test
  decision-making structures}.
\newblock \bibinfo{journal}{\emph{Proceedings - 2010 3rd IEEE International
  Conference on Computer Science and Information Technology, ICCSIT 2010}}
  \bibinfo{volume}{5} (\bibinfo{year}{2010}), \bibinfo{pages}{30--34}.
\newblock
\urldef\tempurl%
\url{https://doi.org/10.1109/ICCSIT.2010.5563989}
\showDOI{\tempurl}
\newblock
\shownote{cited By 10.}


\bibitem[\protect\citeauthoryear{Shahamiri, Wan-Kadir, Ibrahim, and
  Hashim}{Shahamiri et~al\mbox{.}}{2012}]%
        {Shahamiri2012}
\bibfield{author}{\bibinfo{person}{S.R. Shahamiri}, \bibinfo{person}{W.M.N.
  Wan-Kadir}, \bibinfo{person}{S. Ibrahim}, {and} \bibinfo{person}{S.Z.M.
  Hashim}.} \bibinfo{year}{2012}\natexlab{}.
\newblock \showarticletitle{Artificial Neural Networks as multi-networks
  automated test oracle}.
\newblock \bibinfo{journal}{\emph{Automated Software Engineering}}
  \bibinfo{volume}{19}, \bibinfo{number}{3} (\bibinfo{year}{2012}),
  \bibinfo{pages}{303--334}.
\newblock
\urldef\tempurl%
\url{https://doi.org/10.1007/s10515-011-0094-z}
\showDOI{\tempurl}
\newblock
\shownote{cited By 23.}


\bibitem[\protect\citeauthoryear{Singhal, Bansal, and Kumar}{Singhal
  et~al\mbox{.}}{2016}]%
        {Singhal2016}
\bibfield{author}{\bibinfo{person}{A. Singhal}, \bibinfo{person}{A. Bansal},
  {and} \bibinfo{person}{A. Kumar}.} \bibinfo{year}{2016}\natexlab{}.
\newblock \showarticletitle{An approach to design test oracle for aspect
  oriented software systems using soft computing approach}.
\newblock \bibinfo{journal}{\emph{International Journal of Systems Assurance
  Engineering and Management}} \bibinfo{volume}{7}, \bibinfo{number}{1}
  (\bibinfo{year}{2016}), \bibinfo{pages}{1--5}.
\newblock
\urldef\tempurl%
\url{https://doi.org/10.1007/s13198-015-0402-2}
\showDOI{\tempurl}
\newblock
\shownote{cited By 1.}


\bibitem[\protect\citeauthoryear{Taylor and Nitschke}{Taylor and
  Nitschke}{2017}]%
        {taylor2017improving}
\bibfield{author}{\bibinfo{person}{Luke Taylor} {and} \bibinfo{person}{Geoff
  Nitschke}.} \bibinfo{year}{2017}\natexlab{}.
\newblock \bibinfo{title}{Improving Deep Learning using Generic Data
  Augmentation}.
\newblock
\newblock
\showeprint[arxiv]{1708.06020}~[cs.LG]


\bibitem[\protect\citeauthoryear{Vanmali, Last, and Kandel}{Vanmali
  et~al\mbox{.}}{2002}]%
        {Vanmali2002}
\bibfield{author}{\bibinfo{person}{M. Vanmali}, \bibinfo{person}{M. Last},
  {and} \bibinfo{person}{A. Kandel}.} \bibinfo{year}{2002}\natexlab{}.
\newblock \showarticletitle{Using a neural network in the software testing
  process}.
\newblock \bibinfo{journal}{\emph{International Journal of Intelligent
  Systems}} \bibinfo{volume}{17}, \bibinfo{number}{1} (\bibinfo{year}{2002}),
  \bibinfo{pages}{45--62}.
\newblock
\urldef\tempurl%
\url{https://doi.org/10.1002/int.1002}
\showDOI{\tempurl}
\newblock
\shownote{cited By 63.}


\bibitem[\protect\citeauthoryear{{Vineeta}, {Singhal}, and {Bansal}}{{Vineeta}
  et~al\mbox{.}}{2014}]%
        {Vineeta2014a}
\bibfield{author}{\bibinfo{person}{{Vineeta}}, \bibinfo{person}{A. {Singhal}},
  {and} \bibinfo{person}{A. {Bansal}}.} \bibinfo{year}{2014}\natexlab{}.
\newblock \showarticletitle{Generation of test oracles using neural network and
  decision tree model}. In \bibinfo{booktitle}{\emph{2014 5th International
  Conference - Confluence The Next Generation Information Technology Summit
  (Confluence)}}. \bibinfo{pages}{313--318}.
\newblock
\urldef\tempurl%
\url{https://doi.org/10.1109/CONFLUENCE.2014.6949311}
\showDOI{\tempurl}


\bibitem[\protect\citeauthoryear{Ye, Feng, Zhu, and Lin}{Ye
  et~al\mbox{.}}{2006}]%
        {Ye2006a}
\bibfield{author}{\bibinfo{person}{M. Ye}, \bibinfo{person}{B. Feng},
  \bibinfo{person}{L. Zhu}, {and} \bibinfo{person}{Y. Lin}.}
  \bibinfo{year}{2006}\natexlab{}.
\newblock \showarticletitle{Neural networks based automated test oracle for
  software testing}.
\newblock \bibinfo{journal}{\emph{Lecture Notes in Computer Science (including
  subseries Lecture Notes in Artificial Intelligence and Lecture Notes in
  Bioinformatics)}}  \bibinfo{volume}{4234 LNCS - III} (\bibinfo{year}{2006}),
  \bibinfo{pages}{498--507}.
\newblock
\urldef\tempurl%
\url{https://www.scopus.com/inward/record.uri?eid=2-s2.0-33750708220&partnerID=40&md5=858633bde3ccbdac0195bd004eb4141e}
\showURL{%
\tempurl}
\newblock
\shownote{cited By 9.}


\bibitem[\protect\citeauthoryear{{Zhang}, {Zhou}, {Pelliccione}, and
  {Leung}}{{Zhang} et~al\mbox{.}}{2017}]%
        {Zhang2017}
\bibfield{author}{\bibinfo{person}{P. {Zhang}}, \bibinfo{person}{X. {Zhou}},
  \bibinfo{person}{P. {Pelliccione}}, {and} \bibinfo{person}{H. {Leung}}.}
  \bibinfo{year}{2017}\natexlab{}.
\newblock \showarticletitle{RBF-MLMR: A Multi-Label Metamorphic Relation
  Prediction Approach Using RBF Neural Network}.
\newblock \bibinfo{journal}{\emph{IEEE Access}}  \bibinfo{volume}{5}
  (\bibinfo{year}{2017}), \bibinfo{pages}{21791--21805}.
\newblock
\showISSN{2169-3536}
\urldef\tempurl%
\url{https://doi.org/10.1109/ACCESS.2017.2758790}
\showDOI{\tempurl}


\bibitem[\protect\citeauthoryear{Zhang, Wang, and Zhang}{Zhang
  et~al\mbox{.}}{2019}]%
        {Zhang2019d}
\bibfield{author}{\bibinfo{person}{R. Zhang}, \bibinfo{person}{Y.-W. Wang},
  {and} \bibinfo{person}{M.-Z. Zhang}.} \bibinfo{year}{2019}\natexlab{}.
\newblock \showarticletitle{Automatic test oracle based on probabilistic neural
  networks}.
\newblock \bibinfo{journal}{\emph{Advances in Intelligent Systems and
  Computing}}  \bibinfo{volume}{752} (\bibinfo{year}{2019}),
  \bibinfo{pages}{437--445}.
\newblock
\urldef\tempurl%
\url{https://doi.org/10.1007/978-981-10-8944-2_50}
\showDOI{\tempurl}
\newblock
\shownote{cited By 0.}


\end{thebibliography}

\end{document}